\documentclass[aps,pre,twocolumn,superscriptaddress,showpacs]{revtex4}
\usepackage{graphicx}
\usepackage[header,title,page,titletoc]{appendix}

\usepackage{color}
\usepackage{epstopdf}

\begin{document}


\title{The template-specific fidelity of DNA replication with high-order neighbor effects: a first-passage approach}

\author{Qiu-Shi Li}
 \affiliation{School of Physical Science, University of Chinese Academy of Sciences}%

\author{Pei-Dong Zheng}
 \affiliation{School of Physical Science, University of Chinese Academy of Sciences}%

\author{Yao-Gen Shu}%
\affiliation{Institute of Theoretical Physics, Chinese Academy of Sciences}%

\author{Zhong-Can Ou-Yang}%
\affiliation{Institute of Theoretical Physics, Chinese Academy of Sciences}%

\author{Ming Li}
\email{liming@ucas.ac.cn}
\affiliation{School of Physical Science, University of Chinese Academy of Sciences}%
\date{\today}
\begin{abstract}
DNA replication fidelity is a critical issue in molecular biology.
Biochemical experiments have provided key insights on the mechanism of fidelity control by DNAP in the past decades, whereas systematic theoretical studies on this issue began only recently. Because of the underlying difficulties of mathematical treatment, comprehensive surveys on the template-specific replication kinetics are still rare. Here we proposed a first-passage approach to address this problem, in particular the positional fidelity, for complicated processes with high-order neighbor effects. Under biologically-relevant conditions, we derived approximate analytical expressions of the positional fidelity which shows intuitively how some key kinetic pathways are coordinated to guarantee the high fidelity, as well as the high velocity, of the replication processes.
It was also shown that the fidelity at any template position is dominantly determined by the nearest-neighbor template sequences, which is consistent with the idea that replication mutations are randomly distributed in the genome.
\end{abstract}

\maketitle

\section{Introduction}

Since the Watson-Crick (WC) base-pairing rules of double-strand DNA were discovered \cite{watson1953molecular}, template-directed DNA replication has became a critical research subject to understand genetic variations and evolution. It's now widely acknowledged that WC pairings (A-T and G-C, denoted as Right($R$) pairs) play a dominate role in the replication process to maintain the genome stability, while the non-WC pairings (denoted as Wrong($W$) pairs) occur with very low probability (about $10^{-4}$ to $10^{-10}$, dependent on species).
This is not due to the difference between the free energy of $R$ and $W$ pairs in the double-helical DNA: in fact, this free energy difference is only about $2-4 k_{B}T$ which cannot account for such low error rates if estimated by Boltzmann factor.
As pointed out by J.Hopfield\cite{hopfield1974kinetic} and J.Ninio\cite{ninio1975kinetic}, the low error rates originate from the huge difference between the replication kinetics of $R$ and that of $W$, which is realized by high-fidelity DNA polymerases (DNAP) \cite{lehman1958enzymatic,kunkel2000dna}.

DNAP often consists of a polymerase domain and a proofreading domain. The former catalyzes the template-dependent synthesis of the nascent chain. The latter excises the terminal unit of the growing chain, with a higher excision probability for $W$ than for $R$.
While experiments have revealed for a long time that the replication fidelity is determined by both the polymerization kinetics and the proofreading kinetics, related problems were not solved, e.g, how to estimate the positional fidelity (reciprocal of the error rate at each template position), if all the template-specific kinetic parameters are experimentally measured?
Because of the mathematical difficulties of handling the kinetic equations of such complex copolymerization processes, systematic theoretical studies on these issues appeared quite recently. So far there are two categories of models.

One assumes that the kinetic parameters of all $R$ (or $W$) pairs are of the same order of magnitude and thus describes the replication approximately as a $R/W$ binary copolymerization process(i.e, the specific template sequence is not considered explicitly).
This simplification has long been used in biochemistry for theoretical modelling (e.g, see the historical literatures \cite{hopfield1974kinetic,ninio1975kinetic} or more recent publications like \cite{patel1991pre,wong1991induced,tsai2006new,johnson2010kinetic}).
However, thorough studies on such processes appeared only recently, especially for cases in which the rates of monomer addition or deletion at the end of the growing chain depend on the preceding one or more units. Such higher-order neighbor effects may be significant if the terminus of the growing chain contains one or a few $W$s which can destabilize the terminus and hence affect the monomer addition or deletion. These effects have been treated recently by theories under steady-state assumptions, and the overall replication fidelity and growth velocity were calculated numerically or analytically \cite{gaspard2014kinetics,shu2015general,Song2017proofreading}.
In these theories, the copolymerization process was described as a homogenous Markov chain.
This is, however, not appropriate for real cases in which the template DNA sequence is highly inhomogeneous and the kinetic parameters of $R/W$ are highly sequence-dependent.

These template-sequence specificities have not received much attention until very recently. In a series of works, P.Gaspard has considered all the 16 types of base pairs in the kinetic models and handled the high-order neighbor effects successfully \cite{gaspard2016growth,gaspard2016kinetics1,gaspard2016kinetics2,gaspard2016Template,gaspard2017Iterated}
.
By assuming that the probability of any possible sequence of the growing chain can be approximated as a backward (i.e, opposite to the growing direction) inhomogeneous Markov chain in the long-time limit, he succeeded to propose an iteration algorithm to numerically compute the positional fidelity or velocity for any given template sequence (i.e, the fidelity or velocity profile).
However, there are still many questions to be further addressed. For instance,
the iteration algorithm goes through the entire template sequence cyclically for numerical convergence, which indicates that the fidelity at any position may depend on the entire sequence. This is doubtful, for it's a hard to conceive that replication mutations at different positions have long-range correlations rather than randomly distributed as widely believed.
To what a range do the positional quantities depend on the surrounding template sequence? Do the correlations in the template sequence (if any) have any influence on the fidelity or velocity profile?

In this paper, we propose a different approach to address these template-specific problems. Our method is based on a first-passage description of the replication process. This leads to exact expressions of the probability of the nascent chain sequence as forward inhomogenous Markov chains. In contrast to the backward Markov chain assumed in the iteration algorithm\cite{gaspard2017Iterated}, the forward form is more convenient for approximate numerical or analytical calculations which offers intuitive insights on how DNAP achieves high fidelity by proofreading while maintains high velocity. Below we introduce this method, starting from simple binary copolymerization processes with first-order nearest-neighbor effects. We will also show how to generalize this method to more complicated systems.

\section{The basic theory: the first-order replication processes}\label{basic}
For brevity and not losing generality, we suppose that the template sequence consists of two types of units $A$ and $B$, and correspondingly two types of monomers $a$ and $b$ are added to the active end of the growing chain (i.e, the $3'$-end of the nascent DNA chain) and paired with $A$ or $B$ to form a double strand structure. If $a$ pairs with $A$ much more probably than with $B$, we denote $\left(^A_a\right)$ as $R$ and $\left(^A_b\right)$ as $W$. Similarly, we denote $\left(^B_b\right)$ as $R$ and $\left(^B_a\right)$ as $W$.

Given any template sequence of length $L$ (e.g, a region of interests in a real genome), since DNA replication proceeds unidirectionally from the $3'$ end to the $5'$ end of the template,  we assume that the nascent chain initiates from a pre-existing seed (either $a$ or $b$) paired with the $3'$-end unit of the template, then grows and terminates at the $5'$-end of the template.
In the growing stage, the monomer $a$ or $b$ can be added to the end by the polymerase domain of DNAP or deleted from the end by the proofreading domain. In contrast, the initial seed and the lastly-added monomer can not be deleted. In other words, this is a first-passage process from a reflecting boundary at the first position to an absorbing boundary at the last position.
It's worth to note that the initiation and termination here are purely imaginary to simplify the mathematical treatments and do not correspond to the real initiation and termination events in biological DNA replication processes.
We will show later that different choices of the boundary conditions
do not change our major results and conclusions.

For the first-order processes, we assume that the rates of addition or deletion of any monomer $a$ or $b$ depend on the preceding neighbor, denoted as $k^{XY}_{\alpha\beta}$ and $r^{XY}_{\alpha\beta}$ respectively. ($^X_\alpha$) presents the preceding base pair and $Y$ is the template unit to which the monomer $\beta$ is paired, $X, Y = A, B$ and $\alpha, \beta = a , b$.
The termination step occurs with the addition rate of $k^{XY}_{\alpha\beta}$.
It should be noted that all the kinetic parameters here are effective rates. For instance, $k^{XY}_{\alpha\beta}$ is in fact the effective polymerization rate which is contributed by several sub-steps and dependent on the monomer concentrations.
In this manuscript, we will not go into such details.

The probability of the growing chain sequence $\alpha_{1}\alpha_{2}...\alpha_{i} (1\leq i \leq L)$ at time $t$ is denoted as
$p^{X_{1}X_{2}...X_{i}...X_{L}}_{\ \alpha_{1}\alpha_{2}...\alpha_{i}}(t)$. Now we have the following master equations.

\begin{eqnarray}\label{Pmaster}
\dot{p}^{X_{1}...X_{L}}_{\ \alpha_{1}}
&=& r^{X_{1}X_{2}}_{\ \alpha_{1}a}p^{X_{1}X_{2}...X_{L}}_{\ \alpha_{1}a} +r^{X_{1}X_{2}}_{\ \alpha_{1}b}p^{X_{1}X_{2}...X_{L}}_{\ \alpha_{1}b}\nonumber\\
&-&\left(k^{X_{1}X_{2}}_{\ \alpha_{1}a}+k^{X_{1}X_{2}}_{\ \alpha_{1}b}\right) p^{X_{1}...X_{L}}_{\ \alpha_{1}}
\nonumber \\
\dot{p}^{X_{1}...X_{i}...X_{L}}_{\ \alpha_{1}...\alpha_{i}}
&=& k^{X_{i-1}X_{i}}_{\ \alpha_{i-1}\alpha_{i}}p^{X_{1}...X_{i-1}...X_{L}}_{\ \alpha_{1}...\alpha_{i-1}}\nonumber\\
&+&r^{X_{i}X_{i+1}}_{\ \alpha_{i}a}p^{X_{1}...X_{i}X_{i+1}...X_{L}}_{\ \alpha_{1}...\alpha_{i}a} \nonumber \\
&+&r^{X_{i}X_{i+1}}_{\ \alpha_{i}b}p^{X_{1}...X_{i}X_{i+1}...X_{L}}_{\ \alpha_{1}...\alpha_{i}b} \nonumber \\
&-& \left(r^{X_{i-1}X_{i}}_{\ \alpha_{i-1}\alpha_{i}} + k^{X_{i}X_{i+1}}_{\ \alpha_{i}a}+k^{X_{i}X_{i+1}}_{\ \alpha_{i}b}\right)\nonumber\\
&&\cdot p^{X_{1}...X_{i}...X_{L}}_{\ \alpha_{1}...\alpha_{i}}, \ \ 2 \leq i \leq L-2
\nonumber \\
\dot{p}^{X_{1}...X_{L-1}X_{L}}_{\ \alpha_{1}...\alpha_{L-1}}
&=& k^{X_{L-2}X_{L-1}}_{\ \alpha_{L-2}\alpha_{L-1}}p^{X_{1}...X_{L-2}...X_{L}}_{\ \alpha_{1}...\alpha_{L-2}} \nonumber \\
&-& \left(r^{X_{L-2}X_{L-1}}_{\ \alpha_{L-2}\alpha_{L-1}} + k^{X_{L-1}X_{L}}_{\ \alpha_{L-1}a}+k^{X_{L-1}X_{L}}_{\ \alpha_{L-1}b}\right)\nonumber\\
&&\cdot p^{X_{1}...X_{L-1}X_{L}}_{\ \alpha_{1}...\alpha_{L-1}}
\nonumber \\
\dot{p}^{X_{1}...X_{L}}_{\ \alpha_{1}...\alpha_{L}}
&=& k^{X_{L-1}X_{L}}_{\ \alpha_{L-1}\alpha_{L}}p^{X_{1}...X_{L-1}X_{L}}_{\ \alpha_{1}...\alpha_{L-1}}
\end{eqnarray}

One of our major concerns is the final sequence distribution of the nascent chain, i.e, the long-time limit $P^{X_{1}...X_{L}}_{\ \alpha_{1}...\alpha_{L}}=p^{X_{1}...X_{L}}_{\ \alpha_{1}...\alpha_{L}}(t \rightarrow \infty)$. To calculate it, we assume the initial conditions
$p^{X_{1}...X_{L}}_{\ \alpha_{1}}(t=0)= q^{X_{1}}_{\ \alpha_{1}}$, $ q^{X_{1}}_{\ a} + q^{X_{1}}_{\ b} = 1$ ($q^{X_{1}}_{\ \alpha_{1}}$ can be arbitrarily chosen. It has negligible impacts on the fidelity profile except few positions near the reflecting boundary),
$p^{X_{1}...X_{i}...X_{L}}_{\ \alpha_{1}...\alpha_{i}} (t=0) = 0 \ (i \geq 2)$,
and the long-time limits $p^{X_{1}...X_{i}...X_{L}}_{\ \alpha_{1}...\alpha_{i}}(t \rightarrow \infty) = 0 \ (1 \leq i < L)$. We integrate (denoting $\Gamma^{X_{1}...X_{i}...X_{L}}_{\ \alpha_{1}...\alpha_{i}} \equiv  \int^{\infty}_{0} p^{X_{1}...X_{i}...X_{L}}_{\ \alpha_{1}...\alpha_{i}}(t)dt$) and solve the above equations to obtain the following iteration relations

\begin{eqnarray}\label{Pinfty}
&&P^{X_{1}X_{2}...X_{L}}_{\ \alpha_{1}\alpha_{2}...\alpha_{L}} = \left(q^{X_{1}}_{\ \alpha_{1}}/g^{X_{1}...X_{L}}_{\ \alpha_{1}}\right) \cdot
\Pi^{X_{1}X_{2}...X_{L}}_{\ \alpha_{1}\alpha_{2}}\nonumber\\
&&\hspace{1cm}\cdot \Pi^{X_{2}X_{3}...X_{L}}_{\ \alpha_{2}\alpha_{3}}\cdots
\Pi^{X_{L-2}X_{L-1}X_{L}}_{\ \alpha_{L-2}\alpha_{L-1}} \cdot k^{X_{L-1}X_{L}}_{\ \alpha_{L-1}\alpha_{L}}
\\
&&\Pi^{X_{i}X_{i+1}...X_{L}}_{\ \alpha_{i}\alpha_{i+1}} = k^{X_{i}X_{i+1}}_{\ \alpha_{i}\alpha_{i+1}} \Big/ \left(r^{X_{i}X_{i+1}}_{\ \alpha_{i}\alpha_{i+1}} + g^{X_{i+1}...X_{L}}_{\ \alpha_{i+1}}\right)
\nonumber \\
&&g^{X_{i+1}...X_{L}}_{\ \alpha_{i+1}} = \Pi^{X_{i+1}X_{i+2}...X_{L}}_{\ \alpha_{i+1}a} \cdot g^{X_{i+2}...X_{L}}_{\ a}\nonumber\\
&&\hspace{2cm}+ \Pi^{X_{i+1}X_{i+2}...X_{L}}_{\ \alpha_{i+1}b} \cdot g^{X_{i+2}...X_{L}}_{\ b}
\nonumber \\
&&g^{X_{L-1}X_{L}}_{\ \alpha_{L-1}} \equiv k^{X_{L-1}X_{L}}_{\ \alpha_{L-1}a} + k^{X_{L-1}X_{L}}_{\ \alpha_{L-1}b} \nonumber
\end{eqnarray}

Eq.\ref{Pinfty} can be transformed into a more intuitive form, a forward inhomogeneous Markov chain
\begin{eqnarray}\label{Pfinal}
P^{X_{1}X_{2}...X_{L}}_{\ \alpha_{1}\alpha_{2}...\alpha_{L}}&=& q^{X_{1}}_{\ \alpha_{1}} \cdot
M^{X_{1}X_{2}...X_{L}}_{\ \alpha_{1}\alpha_{2}} \cdot M^{X_{2}X_{3}...X_{L}}_{\ \alpha_{2}\alpha_{3}}\nonumber\\
&& \cdots
M^{X_{L-2}X_{L-1}X_{L}}_{\ \alpha_{L-2}\alpha_{L-2}} \cdot M^{X_{L-1}X_{L}}_{\ \alpha_{L-1}\alpha_{L}}
\\
M^{X_{i}X_{i+1}...X_{L}}_{\ \alpha_{i}\alpha_{i+1}} &=& \Pi^{X_{i}X_{i+1}...X_{L}}_{\ \alpha_{i}\alpha_{i+1}}\cdot g^{X_{i+1}...X_{L}}_{\ \alpha_{i+1}} \Big/ g^{X_{i}...X_{L}}_{\ \alpha_{i}}
\nonumber \\
M^{X_{L-1}X_{L}}_{\ \alpha_{L-1}\alpha_{L}} &\equiv& k^{X_{L-1}X_{L}}_{\ \alpha_{L-1}\alpha_{L}} \Big/ g^{X_{L-1}X_{L}}_{\ \alpha_{L-1}}  \nonumber
\end{eqnarray}
Here $M$ is the stochastic transfer matrix with each row sum equals to 1, i.e, $M^{X_{i}X_{i+1}...X_{L}}_{\ \alpha_{i}a} + M^{X_{i}X_{i+1}...X_{L}}_{\ \alpha_{i}b} = 1$. By Eq.\ref{Pfinal}, one can calculate any positional quantities of interest, e.g, the positional probability, $P^{X_m}_{\ \alpha_m} = \sum_{\{\alpha_{i}, i \neq m\}} P^{X_{1}...X_{L}}_{\ \alpha_{1}...\alpha_{L}}$, or equivalently
$ (P^{X_m}_{\ a}, P^{X_m}_{\ b}) = (q^{X_{1}}_{\ a}, q^{X_{1}}_{\ b})\cdot
M^{X_{1}...X_{L}} \cdot \cdot \cdot M^{X_{m-1}X_{m}...X_{L}} $ .

Similarly, we also have
\begin{eqnarray}\label{Gamma}
\Gamma^{X_{1}X_{2}...X_{m}...X_{L}}_{\ \alpha_{1}\alpha_{2}...\alpha_{m}} &=& q^{X_{1}}_{\ \alpha_{1}} \cdot
M^{X_{1}X_{2}...X_{L}}_{\ \alpha_{1}\alpha_{2}} \nonumber\\
&&\cdots
M^{X_{m-1}X_{m}...X_{L}}_{\ \alpha_{m-1}\alpha_{m}} \Big/ g^{X_{m}...X_{L}}_{\ \alpha_{m}}
\end{eqnarray}

\begin{figure*}
\includegraphics[width=12cm]{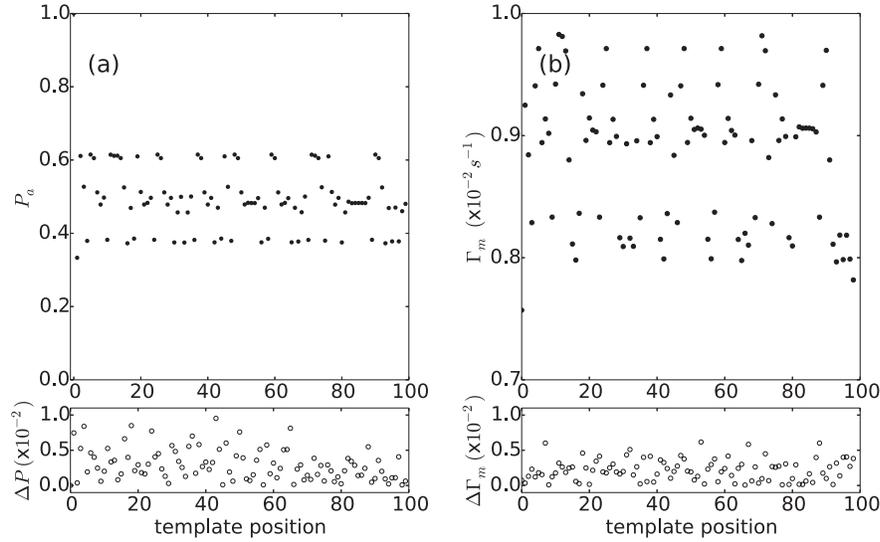}
\caption{The comparison between numerical and simulation results, with given kinetic parameters (Parameters 1, see Appendix \ref{temppara}) and the random template sequence of length 100 (see Appendix \ref{temppara}). The statistics are made over $10^5$ simulations. (a) (top) numerical results of $P_a$ for each template position; (bottom) the relative difference $\Delta P=\max_{\alpha_{m}=a,b}(|P^{num}_{\alpha_{m}}-P^{sim}_{\alpha_{m}}|/P^{num}_{\alpha_{m}})$ . (b) (top)numerical results of the mean dwell time $\Gamma$ for each location; (bottom) the relative difference $\Delta \Gamma_m=|\Gamma_m^{num}-\Gamma_m^{sim}|/\Gamma_m^{num}$ }
\label{f1}
\end{figure*}

Note that the first-passage time (from the position 1 to L) distribution $F(t)$ is determined by the equation $F(t)=-\frac{d}{dt}\sum^{L-1}_{m=1}\sum_{{\{\alpha_{1}...\alpha_{m}\}}}p^{X_{1}X_{2}...X_{m}...X_{L}}_{\ \alpha_{1}\alpha_{2}...\alpha_{m}}(t)$,
it's easy to show that the mean first-passage time $\langle T \rangle = \int^{+\infty}_{0}t\cdot F(t)dt=\sum^{L-1}_{m=1}\Gamma_m $.
Here $\Gamma_m$ is defined as
\begin{eqnarray}\label{Gamma1}
\Gamma_m &=& \sum\limits_{{\{\alpha_{1}...\alpha_{m}\}}}\Gamma^{X_{1}X_{2}...X_{m}...X_{L}}_{\ \alpha_{1}\alpha_{2}...\alpha_{m}}\nonumber\\
&= & \sum\limits_{{\{\alpha_{1}...\alpha_{m}\}}}\int^{+\infty}_{0}p^{X_{1}X_{2}...X_{m}...X_{L}}_{\ \alpha_{1}\alpha_{2}...\alpha_{m}}(t)dt\nonumber\\
&=&\int^{+\infty}_{0}p^{X_{m}}(t)dt
\end{eqnarray}
According to this definition, $\Gamma_m$ is exactly the mean dwell time of the growing chain of length $m$ during the first-passage process (detailed explanations are given in Appendix \ref{dwelltimedis}). In other words, $1/ \Gamma_{m}$ can be regarded as the local growth velocity at position $m$.  $\Gamma_m$ can also be casted in another form
\begin{eqnarray}\label{Gamma2}
\Gamma_m& =& \sum\limits_{{\{\alpha_{1}...\alpha_{m}\}}}\Gamma^{X_{1}X_{2}...X_{m}...X_{L}}_{\ \alpha_{1}\alpha_{2}...\alpha_{m}}\nonumber\\
&=&P^{X_m}_{\ a} \Big/ g^{X_{m}...X_{L}}_{\ a} + P^{X_m}_{\ b} \Big/ g^{X_{m}...X_{L}}_{\ b}
\end{eqnarray}
which is equivalent to Eq.(21)(22) (the mean dwell time at position $m$ calculated by the iteration algorithm) in Ref.\cite{gaspard2017Iterated}, and  $g^{X_{m}...X_{L}}_{\ \alpha}$ is equivalent to $v_{m_l}$ given by Eq.(18) in that paper.

Now the probability profile $P^{X_m}_{a,b}$ and the velocity profile $v_m = 1/\Gamma_m $ can be computed respectively.
Fig.\ref{f1} shows that the numerical results agrees perfectly well with the simulation results given by Gillespie algorithm \cite{gillespie1977exact}.

\begin{figure*}
\includegraphics[width=12cm]{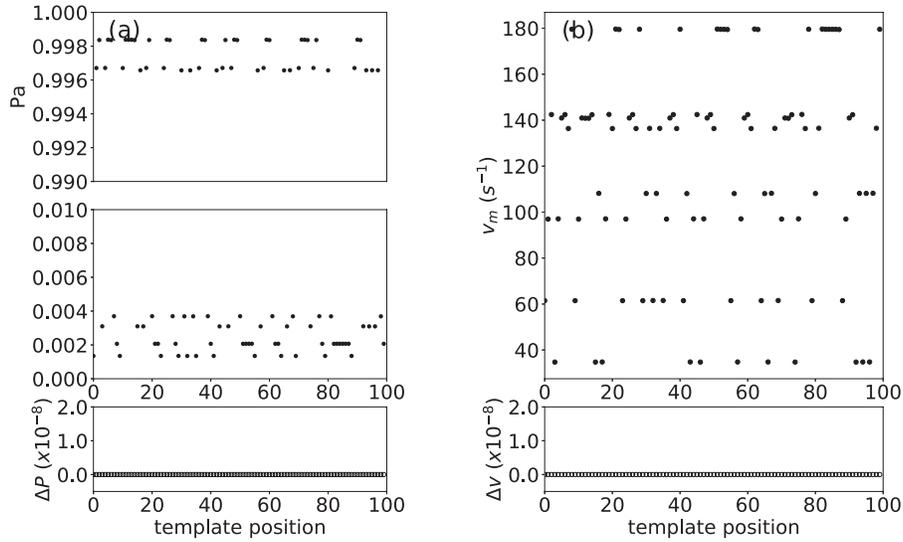}
\caption{Comparison between the numerical results given by FP and IFS under Parameter 2 (Appendix \ref{temppara}). The expanded random template is used in the computations. (a) (top)Pa for each position given by FP algorithm; (bottom) the relative difference $\Delta P=\max_{\alpha_{m}=a,b}|P_{\alpha_{m}}^{FP}-P_{\alpha_{m}}^{IFS}| \big/ P_{\alpha_{m}}^{FP}$. (b) (top) $v_m$ for each position given by FP algorithm; (bottom) the relative difference $\Delta v_m=|v_{\alpha_{m}}^{FP}-v_{\alpha_{m}}^{IFS}|\big/ v_{\alpha_{m}}^{FP}.$}
\label{f2}
\end{figure*}

\begin{figure*}
\includegraphics[width=12cm]{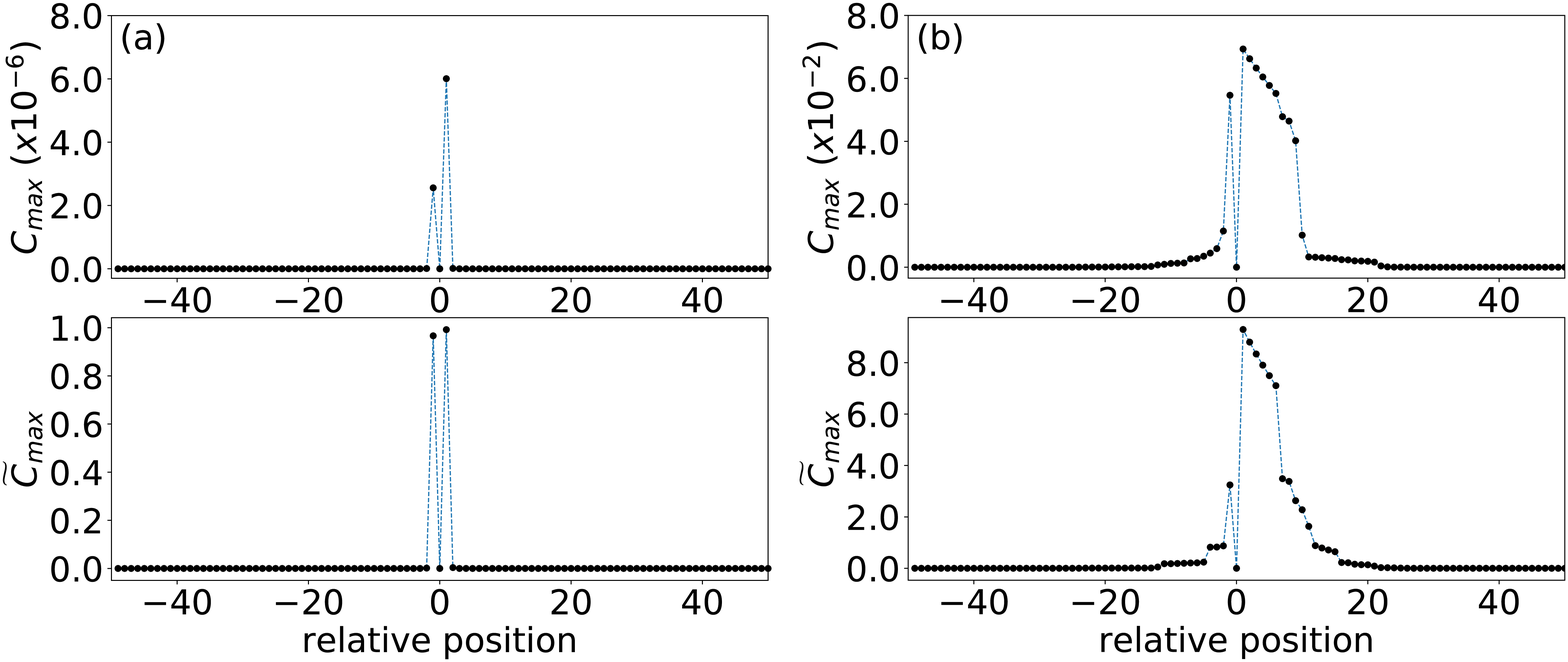}
\caption{The correlation $C_{max}$ and the relative correlation $\widetilde{C}_{max}$ between the position 50 and the rest positions of the random template. (a) under Parameter 2 (bio-relevant conditions). (b) under Parameter 3.}
\label{f3}
\end{figure*}

Our first-passage (FP) calculations are also in perfect agreement with the numerical results given by the iteration algorithm (denoted as IFS in Ref. \cite{gaspard2017Iterated}), as shown by the illustrative example in Fig.\ref{f2}.
It should be point out that since the two algorithm assume different boundary conditions, the numerical results are somewhat different near the two boundaries.
However, by expanding the template sequence from both ends in our FP algorithm,
the difference can be largely decreased or even eliminated.
For instance, the original template sequence is repeated three times to get an expanded new template, and the computed profiles of the middle copy shows no difference with the results of IFS algorithm (Fig.\ref{f2}). This treatment and the expanded template are also used to obtain Fig.\ref{f3}, Fig.\ref{f4} and Fig.\ref{f5}.

\section{Correlations in the probability profile}\label{correlation}

Correlations could be present in the probability profile due to the nearest or higher-order neighbor effects. To see if there are long-range correlations in the first-order processes, we calculate the correlation function between any two template positions, say $i,j$. The function is defined as $C^{X_i.X_j}_{\ \alpha_i.\alpha_j}=\sum_{\{\alpha_k, k\neq i,j\}}P^{X_1..X_k..X_L}_{\ \alpha_1..\alpha_k..\alpha_L} - P^{X_i}_{\ \alpha_i} \cdot P^{X_j}_{\ \alpha_j}$.
There are four types of correlation functions. To quantify the maximal correlations, we define $C_{max}(i,d)=\max_{\alpha_{i},\alpha_{i+d}}(|C^{X_{i},X_{i+d}}_{\ \alpha_{i},\alpha_{i+d}}|)$, $C_{max}(i,0)=0$, for any position $i$,
and correspondingly the relative correlation function $\widetilde{C}^{X_i.X_j}_{\ \alpha_i.\alpha_j}=C^{X_i.X_j}_{\ \alpha_i.\alpha_j} \Big/ \left(P^{X_i}_{\ \alpha_i} \cdot P^{X_j}_{\ \alpha_j}\right)$ and $\widetilde{C}_{max}(i,d)=\max_{\alpha_{i},\alpha_{i+d}}(|\widetilde{C}^{X_{i},X_{i+d}}_{\ \alpha_{i},\alpha_{i+d}}|)$.

Under some conditions (e.g, the bio-relevant conditions such as Parameters 2, which is explained in details in Section \ref{NNapp}), either $C_{max}(i,d)$ or $\widetilde{C}_{max}(i,d)$ decays abruptly with the correlation length 1 (illustrated by Fig.\ref{f3}(a)), implying that the positional probability is determined by its nearest neighbors.
This does not hold in general, of course. For instance, the correlation length will become much larger under some extreme conditions (e.g, Parameter 3 in Fig.\ref{f3}(b)) where one can not identify consistent pairing rules for each type of template units. For instance, $\left(^A_a\right)$ is the dominant pairing (say, with a probability larger than 0.9) only for a part of the template $A$s while $\left(^A_b\right)$ is dominant for the rest, so no Watson-Crick like pairing rules ($R$ or $W$) can be universally assigned to $A$.
Therefore, DNA synthesized in these cases can no longer fulfil its fundamental role as  genetic material.
Such extreme conditions and long-range correlations are out of the scope of this manuscript and will not be discussed in later sections.

\section{The nearest-neighbor approximation under bio-relevant conditions}\label{NNapp}

The nearest neighbor correlations can be observed under the so-called biologically-relevant conditions which are inspired by the measured kinetic parameters of real DNAPs. These conditions ensure that, compared with the replication catalyzed only by the polymerase domain of DNAP,  the introduction of proofreading domain can significantly enhance the replication fidelity while still maintain the high overall velocity.

The bio-relevant conditions for the first-order process are intuitive, as below.

(a)$k^{XY}_{RR}\gg k^{XY}_{RW}$, which mean that the addition of $R$ are always much faster than that of $W$.

(b)$k^{XY}_{WR}\Big/k^{XY}_{WW}\gg k^{XY}_{RR}\Big/k^{XY}_{RW}$, which mean that the successive additions of $W$ are almost inhibited. In fact, $k^{XY}_{WW}$ are hard to measure in experiments, so $k^{XY}_{WW}\sim 0$ are always assumed.

(c)$k^{YZ}_{RR}\gg r^{XY}_{RR},r^{XY}_{WR}$, which mean that the successive additions of $R$ always dominate the replication process in order to guarantee the high replication velocity  (i.e, the introduction of proofreading almost does not decrease the overall velocity), at the cost of that a buried $W$ is hard to be proofread.

(d)$r^{XY}_{WW}> r^{XY}_{RW}$, which mean that the terminus containing more $W$s is more likely to be proofread.

Here $ \gg $ means that the term on the left side is more than 10 times bigger than that on the right side.
These conditions are consistent with experimental observations of real DNAPs (see Sec.3.2 in Ref. \cite{Song2017proofreading} for the data), and in fact are much more general (for comparison, e.g, $k^{XY}_{RR}/k^{XY}_{RW} > 10^5$ and  $k^{XY}_{RR} >> k^{XY}_{WR}$ are always observed in real DNAPs).
Under such general conditions, the exact method introduced in Sec.\ref{basic} can be well approximated by the following method. We start from the iteration relations

\begin{eqnarray}\label{}
g^{X_{i}...X_{L}}_{\ \alpha_{i}} &=& \frac{k^{X_{i}X_{i+1}}_{\ \alpha_{i} a}}{1+ r^{X_{i}X_{i+1}}_{\ \alpha_{i} a} \big/ g^{X_{i+1}...X_{L}}_{\ a}}\nonumber\\
&+& \frac{k^{X_{i}X_{i+1}}_{\ \alpha_{i} b}}{1+ r^{X_{i}X_{i+1}}_{\ \alpha_{i} b} \big/ g^{X_{i+1}...X_{L}}_{\ b}}
\\
g^{X_{L-1}X_{L}}_{\ \alpha_{L-1}} &\equiv& k^{X_{L-1}X_{L}}_{\ \alpha_{L-1}a} + k^{X_{L-1}X_{L}}_{\ \alpha_{L-1}b} \nonumber
\end{eqnarray}

Under bio-relevant conditions, one has $k^{X_{L-1}X_{L}}_{\ \alpha_{L-1} R} >> k^{X_{L-1}X_{L}}_{\ \alpha_{L-1} W}$, so $g^{X_{L-1}X_{L}}_{\ \alpha_{L-1}} \simeq k^{X_{L-1}X_{L}}_{\ \alpha_{L-1} R} $.

The next iteration is

\begin{eqnarray}\label{}
g^{X_{L-2}X_{L-1}X_{L}}_{\ \alpha_{L-2}} &=& \frac{k^{X_{L-2}X_{L-1}}_{\ \alpha_{L-2} a}}{1+ r^{X_{L-2}X_{L-1}}_{\ \alpha_{L-2} a} \big/ g^{X_{L-1}X_{L}}_{\ a}}\nonumber\\
  &+& \frac{k^{X_{L-2}X_{L-1}}_{\ \alpha_{L-2} b}}{1+ r^{X_{L-2}X_{L-11}}_{\ \alpha_{L-2} b} \big/ g^{X_{L-1}X_{L}}_{\ b}}
\end{eqnarray}

If $X_{L-1}=A$, then $r^{X_{L-2}A}_{\ \alpha_{L-2} a} << k^{A X_{L}}_{aR} \simeq g^{A X_{L}}_{a}$ and
$k^{X_{L-2}A}_{\ \alpha_{L-2} a} >> k^{X_{L-2}A}_{\ \alpha_{L-2} b}$. This leads to
$g^{X_{L-2}AX_{L}}_{\ \alpha_{L-2}} \simeq k^{X_{L-2}A}_{\ \alpha_{L-2} a} $.
If $X_{L-1}=B$, then $r^{X_{L-2}B}_{\ \alpha_{L-2} b} << k^{B X_{L}}_{bR} \simeq g^{B X_{L}}_{b}$  and
$k^{X_{L-2}B}_{\ \alpha_{L-2} b} >> k^{X_{L-2}B}_{\ \alpha_{L-2} a}$. This leads to
$g^{X_{L-2}BX_{L}}_{\ \alpha_{L-2}} \simeq k^{X_{L-2}B}_{\ \alpha_{L-2} b} $.
Combining these two results, we get $g^{X_{L-2}X_{L-1}X_{L}}_{\ \alpha_{L-2}} \simeq k^{X_{L-2}X_{L-1}}_{\ \alpha_{L-2} R}$.

Following the same logic, we obtain $g^{X_{i}X_{i+1}...X_{L}}_{\ \alpha_{i}} \simeq k^{X_{i}X_{i+1}}_{\ \alpha_{i} R}$ (denoted as $g^{X_{i}X_{i+1}}_{\ \alpha_{i}}$)
and $\prod^{X_{i}X_{i+1}...X_{L}}_{\ \alpha_{i}\alpha_{i+1}} \simeq k^{X_{i}X_{i+1}}_{\ \alpha_{i}\alpha_{i+1}} / (r^{X_{i}X_{i+1}}_{\ \alpha_{i}\alpha_{i+1}} + k^{X_{i+1}X_{i+2}}_{\ \alpha_{i+1}R}) $
(denoted as $\prod^{X_{i}X_{i+1}X_{i+2}}_{\ \alpha_{i}\alpha_{i+1}}$).
Correspondingly, the stochastic transfer matrix is approximated as $M^{X_{i}X_{i+1}X_{i+2}}_{\ \alpha_{i}\alpha_{i+1}}$ (the nearest-neighbor or NN approximation) which can be transformed by row or column exchange into the equivalent form
\begin{displaymath}
\left[ \begin{array}{cc}
M_{RR} & M_{RW} \\
M_{WR} & M_{WW}
\end{array} \right]
\end{displaymath}
and correspondingly,
\begin{eqnarray}
P^{X_{1}X_{2}...X_{L}}_{\ s_{1}s_{2}...s_{L}} &\simeq& q^{X_{1}}_{\ s_{1}} \cdot
M^{X_{1}X_{2}X_{3}}_{\ s_{1}s_{2}} \cdots M^{X_{i-1}X_{i}X_{i+1}}_{\ s_{i-1}s_{i}} \nonumber\\
&&\cdots M^{X_{L-1}X_{L}}_{\ s_{L-1}s_{L}}, \nonumber\\
\ s_i &=& R,W \ (i=1,...,L)
\end{eqnarray}

\begin{figure*}
\includegraphics[width=12cm]{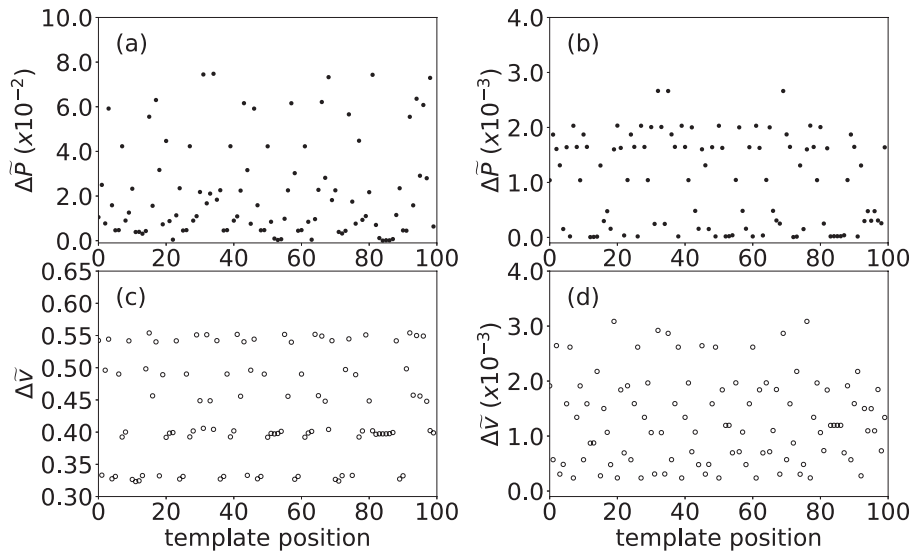}
\caption{Comparison between the precise (pre) and approximate (app) numerical results. $\Delta{\widetilde{P}_{i}}=\max_{\alpha_i=a,b}\left(|P^{app}_{\alpha_i}-P^{pre}_{\alpha_i}|\big/P^{pre}_{\alpha_i}\right)$ and $\Delta{\widetilde{v}}=|v^{app}-v^{pre}|\big/v^{pre}$. (a)(c) $\Delta{\widetilde{P}}$, $\Delta{\widetilde{v}}$, under Parameter 1. (b)(d) $\Delta{\widetilde{P}}$, $\Delta{\widetilde{v}}$, under Parameter 2.}
\label{f4}
\end{figure*}

Now we get the approximate expressions of the elements of $M$. For instance,
\begin{eqnarray}\label{Mrw}
M^{X_{i}X_{i+1}X_{i+2}}_{RW} &\simeq & \frac{k^{X_{i}X_{i+1}}_{RW}}{r^{X_{i}X_{i+1}}_{RW} + g^{X_{i+1}X_{i+2}}_{W}} \frac{g^{X_{i+1}X_{i+2}}_{W}}{g^{X_{i}X_{i+1}}_{R}} \nonumber \\
&\simeq & \frac{k^{X_{i}X_{i+1}}_{RW}}{r^{X_{i}X_{i+1}}_{RW} + k^{X_{i+1}X_{i+2}}_{WR}} \frac{k^{X_{i+1}X_{i+2}}_{WR}}{k^{X_{i}X_{i+1}}_{RR}}  \nonumber \\
&=& \frac{k^{X_{i}X_{i+1}}_{RW}}{k^{X_{i}X_{i+1}}_{RR}} \Bigg/ \left(1+ \frac{r^{X_{i}X_{i+1}}_{RW}}{k^{X_{i+1}X_{i+2}}_{WR}}\right)
\end{eqnarray}

It can be shown that $M_{RR} >> M_{RW}$, $M_{WR} >> M_{WW} $ and $M_{RW} >> M_{WW}$ always hold under bio-relevant conditions. For stochastic matrices like $M$

\begin{displaymath}
\left[ \begin{array}{cc}
1-y & y \\
1-z & z
\end{array} \right]
\end{displaymath}
with $z << y << 1$, one can verify that its left eigenvector associated with the largest eigenvalue 1 is approximately $ P=(1-y,y)$ ($\lim\limits_{n\rightarrow \infty} M^n$ converges to a matrix in which each row is the eigenvector $P$. For more details of heuristic analysis, see Appendix \ref{eigenapp}.).
$P$ is a good approximation of the precise probability distribution at position $i+1$, which can be verified numerically (see Figs.\ref{f4}(b,d)). Even under some conditions different from bio-relevant conditions (Parameters 1), the NN approximation can also give results of the same orders of magnitude with the precise numerical results (Figs.\ref{f4}(a,c)). This approximation, however, fails under conditions far different from bio-relevant conditions (data not shown here. see Supplementary Materials for more details).

\begin{figure*}
\includegraphics[width=12cm]{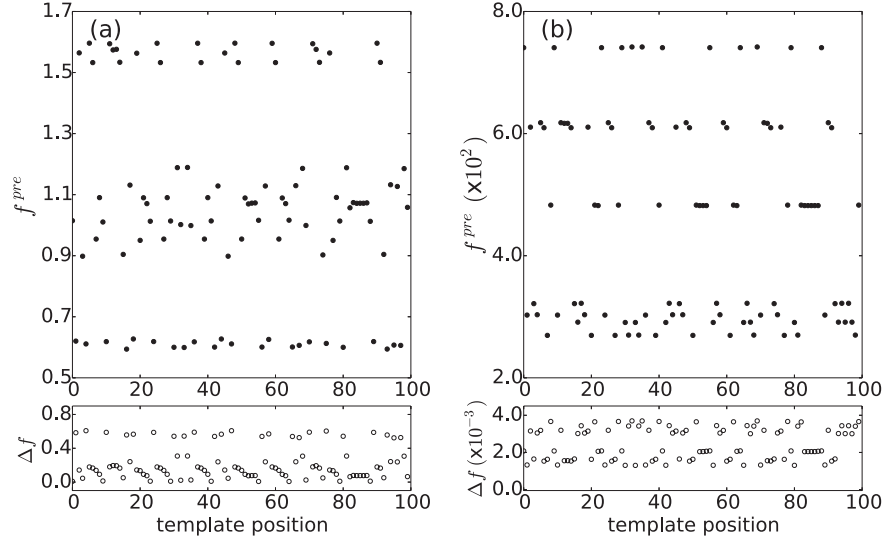}
\caption{Comparison between the precise (pre) and approximated (app) fidelity profile.  $\Delta f=|f^{pre}-f^{app}|/f^{pre}$. (a)$f^{pre}$ (top) and $\Delta f$ (bottom), under Parameter 1. (b)$f^{pre}$ (top) and $\Delta f$ (bottom),under Parameter 2.}
\label{f5}
\end{figure*}

One can also obtain the analytical expressions of the probability profile $(P^{i+1}_R,P^{i+1}_W) \simeq (M^{X_{i}X_{i+1}X_{i+2}}_{RR}, M^{X_{i}X_{i+1}X_{i+2}}_{RW})$ and thus gives the approximate fidelity profile

\begin{eqnarray}\label{}
f^{X_{i+1}}\simeq \frac{M^{X_{i}X_{i+1}X_{i+2}}_{RR}}{M^{X_{i}X_{i+1}X_{i+2}}_{RW}} \simeq
 \frac{k^{X_{i}X_{i+1}}_{RR}}{k^{X_{i}X_{i+1}}_{RW}}\left(1+ \frac{r^{X_{i}X_{i+1}}_{RW}}{k^{X_{i+1}X_{i+2}}_{WR}}\right)
\end{eqnarray}
which includes the analytical expression of fidelity Eq.(15) in Ref.\cite{Song2017proofreading} (template-sequence specificity ignored), as a limiting case.
The approximate profile shows perfect agreement with the precise profile under bio-relevant conditions (Fig.\ref{f5}(b)), and also provides a good estimate under some other conditions (Fig.\ref{f5}(a) where the approximate and the precise velocity is of the same order of magnitude).

The NN approximation immediately leads to the conclusion that any kind of correlations in the template sequence, if exists (e.g, the possible long-range correlations in the non-coding DNA sequences \cite{LiWT2007}), has no substantial impact on the positional quantities (data are shown in Supplementary Materials).
This is consistent with the widely acknowledged idea that DNA replication mutations are randomly distributed in the genome.

\section{Generalization to multi-component systems} \label{multicomp}
The above methods can be readily generalized to more realistic cases, e.g, in real DNA replication there are four types of monomers (A,G,T,C) being added or deleted.
Below we consider a general multi-component system which consist of $n$ types of units $A_i \ (i=1,..,n)$ in the template and $n$ types of monomers $a_i \ (i=1,..,n)$ in the solution, and each $A_i$ forms the right pair (R) with only one monomer $a_i $ and forms wrong pairs with the rest monomers (denoted as W$_i, i=1,2,...,n-1$).
The corresponding bio-relevant conditions are just a simple generalization of those in the preceding section.

(a)$k^{XY}_{RR}\gg k^{XY}_{RW_{i}}$

(b)$k^{XY}_{W_{i}R}\Big/k^{XY}_{W_{i}W_{j}}\gg k^{XY}_{RR}\Big/k^{XY}_{RW_{k}}$

(c)$k^{YZ}_{RR}\gg r^{XY}_{RR},r^{XY}_{W_{i}R}$

(d) $r^{XY}_{W_{i}W_{j}}>r^{XY}_{RW_{j}}$

Similarly we rearrange the transfer matrix $M^{X_{i-1}X_{i}X_{i+2}}$ to a standard form
\begin{displaymath}
\left[ \begin{array}{cccc}
M_{RR} & M_{RW_1} & \ldots & M_{RW_{(n-1)}} \\
M^{(1)}_{WR} & M^{(1)}_{WW_1} & \ldots & M^{(1)}_{WW_{(n-1)}} \\
 & &\ldots&\\
M^{(n-1)}_{WR} & M^{(n-1)}_{WW_1} & \ldots & M^{(n-1)}_{WW_{(n-1)}}
\end{array} \right]
\end{displaymath}
It can be shown that $M_{WW}<<M_{RW}<<1$ in general under bio-relevant conditions,
so the eigenvector $V_1$ of this matrix is approximately $(M_{RR}, M_{RW_1}, \cdot\cdot\cdot, M_{RW_{(n-1)}})$ which is a good approximation of the probability vector $P^{X_{i}}$.

Simple calculations give results almost the same as Eq.\ref{Mrw}, i.e,
\begin{eqnarray}\label{}
M^{X_{i-1}X_{i}X_{i+1}}_{RW_j} &=& \frac{k^{X_{i-1}X_{i}}_{RW_j}}{k^{X_{i-1}X_{i}}_{RR}} \Bigg/ \left(1+ \frac{r^{X_{i-1}X_{i}}_{RW_j}}{k^{X_{i}X_{i+1}}_{W_jR}}\right),
\end{eqnarray}
where $j=1,...,n-1$.

Now the positional fidelity at $i$ is $ f^{X_{i}} = M^{X_{i-1}X_{i}X_{i+1}}_{RR} \Big/ \left(\sum^{n-1}_{j=1} M^{X_{i-1}X_{i}X_{i+1}}_{RW_j}\right) $.

\section{Generalization to higher order processes} \label{higher}
For $h$-order processes, we set the initial seed as a given distribution $q^{X_{1}...X_{h}}_{\ \alpha_{1}...\alpha_{h}}$.
One can follow the logic of Sec.\ref{basic} to obtain
\begin{eqnarray}\label{}
P^{X_{1}...X_{L}}_{\ \alpha_{1}...\alpha_{L}} &=& \left(q^{X_{1}...X_{h}}_{\ \alpha_{1}...\alpha_{h}} \Big/g^{X_{1}...X_{h}...X_{L}}_{\ \alpha_{1}...\alpha_{h}} \right)\nonumber\\
&& \cdot
\Pi^{X_{1}...X_{h+1}...X_{L}}_{\ \alpha_{1}...\alpha_{h+1}} \cdots \Pi^{X_{i}...X_{i+h}...X_{L}}_{\ \alpha_{i}...\alpha_{i+h}}  \nonumber \\
&& \cdots
\Pi^{X_{L-h-1}...X_{L-1}X_{L}}_{\ \alpha_{L-h-1}...\alpha_{L-1}} \cdot k^{X_{L-h}...X_{L}}_{\ \alpha_{L-h}...\alpha_{L}}\nonumber\\
\Pi^{X_{i}...X_{i+h}...X_{L}}_{\ \alpha_{i}...\alpha_{i+h}} &=& k^{X_{i}...X_{i+h}}_{\ \alpha_{i}...\alpha_{i+h}} \Big/\nonumber\\
&& \left(r^{X_{i}...X_{i+h}}_{\ \alpha_{i}...\alpha_{i+h}} + g^{X_{i+1}...X_{i+h}...X_{L}}_{\ \alpha_{i+1}...\alpha_{i+h}}\right)
\nonumber \\
g^{X_{i+1}...X_{i+h}...X_{L}}_{\ \alpha_{i+1}...\alpha_{i+h}} &=& \Pi^{X_{i+1}...X_{i+h}X_{i+h+1}...X_{L}}_{\ \alpha_{i+1}...\alpha_{i+h}a}\nonumber\\
&& \cdot g^{X_{i+2}...X_{i+h}X_{i+h+1}...X_{L}}_{\ \alpha_{i+2}...\alpha_{i+h}a}
 \nonumber \\
& +& \Pi^{X_{i+1}...X_{i+h}X_{i+h+1}...X_{L}}_{\ \alpha_{i+1}...\alpha_{i+h}b} \nonumber\\
&&\cdot g^{X_{i+2}...X_{i+h}X_{i+h+1}...X_{L}}_{\ \alpha_{i+2}...\alpha_{i+h}b}
\nonumber \\
g^{X_{L-h}...X_{L-1}X_{L}}_{\ \alpha_{L-h}...\alpha_{L-1}} &\equiv& k^{X_{L-h}...X_{L}}_{\ \alpha_{L-h}...a} + k^{X_{L-h}...X_{L}}_{\ \alpha_{L-h}...b}
\end{eqnarray}

or equivalently,
\begin{eqnarray}\label{}
P^{X_{1}X_{2}...X_{L}}_{\ \alpha_{1}\alpha_{2}...\alpha_{L}}&=& q^{X_{1}...X_{h}}_{\ \alpha_{1}...\alpha_{h}} \cdot
M^{X_{1}...X_{h+1}...X_{L}}_{\ \alpha_{1}...\alpha_{h+1}} \nonumber\\
&\cdots& M^{X_{i}...X_{i+h}...X_{L}}_{\ \alpha_{i}...\alpha_{i+h}} \cdots
M^{X_{L-h}...X_{L}}_{\ \alpha_{L-h}...\alpha_{L}}\nonumber\\
M^{X_{i}...X_{i+h}...X_{L}}_{\ \alpha_{i}...\alpha_{i+h}}
&=& \Pi^{X_{i}...X_{i+h}...X_{L}}_{\ \alpha_{i}...\alpha_{i+h}} \cdot g^{X_{i+1}...X_{i+h}...X_{L}}_{\ \alpha_{i+1}...\alpha_{i+h}} \nonumber\\ &&\Big/g^{X_{i}...X_{i+h-1}...X_{L}}_{\ \alpha_{i}...\alpha_{i+h-1}}
\nonumber \\
M^{X_{L-h}...X_{L}}_{\ \alpha_{L-h}...\alpha_{L}} &\equiv& k^{X_{L-h}...X_{L}}_{\ \alpha_{L-h}...\alpha_{L}} \Big/ g^{X_{L-h}...X_{L-1}X_{L}}_{\ \alpha_{L-h}...\alpha_{L-1}}
\end{eqnarray}

The NN approximations can also applied to these processes under the corresponding  bio-relevant conditions. For illustration, we only give a brief introduction to the second-order processes of binary systems. The bio-relevant conditions similar to that in Sec.\ref{NNapp} are proposed as below.

(a)$k^{XYZ}_{\alpha\beta R}\gg k^{XYZ}_{\alpha\beta W}$, $\alpha,\beta=R,W$, which mean that the addition rates of $R$ are always much larger than that of $W$.

(b)$\widetilde{k}^{XYZ}_{\alpha\beta R}\Big/\widetilde{k}^{XYZ}_{\alpha\beta W}\gg\widetilde{k}^{XYZ}_{RRR}\Big/\widetilde{k}^{XYZ}_{RRW}$, $\alpha\beta=RW,WR,WW$.
$\tilde{k}^{XYZ}_{\alpha\beta\gamma}\equiv k^{XYZ}_{\alpha\beta\gamma}\Big/\Big(1+r^{XYZ}_{\alpha\beta\gamma}\Big/k^{YZU}_{\beta\gamma R}\Big)$ is approximately the renormalized addition rates.
In fact, here $k^{XYZ}_{\alpha\beta W}\simeq 0$ are always assumed since successive additions of $W$ are almost inhibited. So these conditions are naturally satisfied.

(c)$k^{YZU}_{RRR}\gg r^{XYZ}_{RRR},r^{XYZ}_{WRR}$, which mean that the successive additions of $R$ always dominate the overall replication process.

(d)$r^{XYZ}_{WWR}>r^{XYZ}_{RWR}$, which mean that the terminus containing more $W$s is more likely to be proofread.

(e)$r^{XYZ}_{\alpha\beta W}\Big/k^{YZU}_{\beta WR}>r^{XYZ}_{\alpha\beta R}\Big/k^{YZU}_{\beta RR}$, $\alpha,\beta=R,W$, which mean that the terminal $W$ is always more probable to be deleted than the terminal $R$.

To calculate the positional quantities at position $i$, we first obtain the transfer matrix $M^{X_{i-2}X_{i-1}X_{i}X_{i+1}X_{i+2}}$ by the following two iterations, starting from $g^{X_{i}X_{i+1}X_{i+2}}_{\ \alpha_i \alpha_{i+1}} \simeq k^{X_{i}X_{i+1}X_{i+2}}_{\ \alpha_i \alpha_{i+1} a} + k^{X_{i}X_{i+1}X_{i+2}}_{\ \alpha_i \alpha_{i+1} b} \simeq k^{X_{i}X_{i+1}X_{i+2}}_{\ \alpha_i \alpha_{i+1} R} $,

\begin{eqnarray}\label{}
&&\textbf{(I)}\ \Pi^{X_{i-1}X_{i}X_{i+1}X_{i+2}}_{\ \alpha_{i-1} \alpha_i \alpha_{i+1}} = k^{X_{i-1}X_{i}X_{i+1}}_{\ \alpha_{i-1} \alpha_{i} \alpha_{i+1}} \Big/ \nonumber\\
&&\hspace{3cm} \left(r^{X_{i-1}X_{i}X_{i+1}}_{\ \alpha_{i-1} \alpha_{i} \alpha_{i+1}}+ g^{X_{i}X_{i+1}X_{i+2}}_{\ \alpha_i \alpha_{i+1}}\right)
\nonumber \\
&&g^{X_{i-1}X_{i}X_{i+1}X_{i+2}}_{\ \alpha_{i-1} \alpha_i} = \Pi^{X_{i-1}X_{i}X_{i+1}X_{i+2}}_{\ \alpha_{i-1} \alpha_i a} g^{X_{i}X_{i+1}X_{i+2}}_{\ \alpha_i a}\nonumber\\
&&\hspace{3cm} + \Pi^{X_{i-1}X_{i}X_{i+1}X_{i+2}}_{\ \alpha_{i-1} \alpha_i b} g^{X_{i}X_{i+1}X_{i+2}}_{\ \alpha_i b}
\nonumber \\
&&\textbf{(II)}\ \Pi^{X_{i-2}X_{i-1}X_{i}X_{i+1}X_{i+2}}_{\ \alpha_{i-2} \alpha_{i-1} \alpha_{i}} = k^{X_{i-2}X_{i-1}X_{i}}_{\ \alpha_{i-2} \alpha_{i-1} \alpha_{i}} \Big/\nonumber\\
&&\hspace{2cm}\left(r^{X_{i-2}X_{i-1}X_{i}}_{\ \alpha_{i-2} \alpha_{i-1} \alpha_{i}}+ g^{X_{i-1}X_{i}X_{i+1}X_{i+2}}_{\ \alpha_{i-1} \alpha_{i}}\right)
\nonumber \\
 &&g^{X_{i-2}X_{i-1}X_{i}X_{i+1}X_{i+2}}_{\ \alpha_{i-2} \alpha_{i-1}} = \Pi^{X_{i-2}X_{i-1}X_{i}X_{i+1}X_{i+2}}_{\ \alpha_{i-2} \alpha_{i-1} a} \nonumber\\
 &&\hspace{5cm}\cdot g^{X_{i-1}X_{i}X_{i+1}X_{i+2}}_{\ \alpha_{i-1} a}\nonumber\\
  &&\hspace{2cm}+ \Pi^{X_{i-2}X_{i-1}X_{i}X_{i+1}X_{i+2}}_{\ \alpha_{i-2} \alpha_{i-1} a}
 g^{X_{i-1}X_{i}X_{i+1}X_{i+2}}_{\ \alpha_{i-1} b}
\nonumber\\
&&M^{X_{i-2}X_{i-1}X_{i}X_{i+1}X_{i+2}}_{\ \alpha_{i-2} \alpha_{i-1} \alpha_{i}} = \Pi^{X_{i-2}X_{i-1}X_{i}X_{i+1}X_{i+2}}_{\ \alpha_{i-2} \alpha_{i-1} \alpha_{i}}\nonumber\\
&&\hspace{2cm}g^{X_{i-1}X_{i}X_{i+1}X_{i+2}}_{\ \alpha_{i-1} \alpha_i} \Big/ g^{X_{i-2}X_{i-1}X_{i}X_{i+1}X_{i+2}}_{\ \alpha_{i-2} \alpha_{i-1}}\nonumber
\end{eqnarray}
$M$ can be rewritten as first-order Markov transfer matrix, with four rows indexed as $^{X_{i-2}X_{i-1}}_{s_{i-2}s_{i-1}}$ ($RR,RW,WR,WW$ from up to bottom) and four columns indexed as $^{X_{i-1}X_{i}}_{s_{i-1}s_{i}}$ ($RR,RW,WR,WW$ from left to right) , $s=R,W$
\begin{displaymath}
\left[ \begin{array}{cccc}
M_{RRR} & M_{RRW} & 0 & 0 \\
0 & 0 & M_{RWR} & M_{RWW} \\
 M_{WRR} & M_{WRW}&0&0\\
0 & 0 & M_{WWR} & M_{WWW}
\end{array} \right]
\end{displaymath}

It can be shown that $M_{WWW}, M_{RWW}, M_{WRW} << M_{RRW} << 1$. The
eigenvector $V_1$ of this matrix is approximately $(1-2M_{RRW}, M_{RRW}, M_{RRW},0)$ which can be regarded as the positional probability $P^{X_{i-1}X_{i}}=(P^{X_{i-1}X_{i}}_{RR},P^{X_{i-1}X_{i}}_{RW},P^{X_{i-1}X_{i}}_{WR}, P^{X_{i-1}X_{i}}_{WW})$.

To be specific,
\begin{eqnarray}\label{}
&&M^{X_{i-2}X_{i-1}X_{i}X_{i+1}X_{i+2}}_{R_{i-2}R_{i-1}W_{i}} \simeq \frac{k^{X_{i-2}X_{i-1}X_{i}}_{R_{i-2}R_{i-1}W_{i}}}{k^{X_{i-2}X_{i-1}X_{i}}_{R_{i-2}R_{i-1}R_{i}}}\Bigg{/}\nonumber\\
&&\hspace{1cm}\left[1+ \frac{r^{X_{i-2}X_{i-1}X_{i}}_{R_{i-2}R_{i-1}W_{i}}}{k^{X_{i-1}X_{i}X_{i+1}}_{R_{i-1}W_{i}R_{i+1}}}
\left(1+ \frac{r^{X_{i-1}X_{i}X_{i+1}}_{R_{i-1}W_{i}R_{i+1}}}{k^{X_{i}X_{i+1}X_{i+2}}_{W_{i}R_{i+1}R_{i+2}}}\right)\right]
\end{eqnarray}

The positional probability at position $i$ can be calculated by $P^{X_i}_{R} = P^{X_{i-1}X_{i}}_{RR} + P^{X_{i-1}X_{i}}_{WR} = 1-M_{RRW} = M_{RRR} (\simeq 1)$, $P^{X_i}_{W} = P^{X_{i-1}X_{i}}_{RW} + P^{X_{i-1}X_{i}}_{WW} = M_{RRW} $.
So the fidelity at position $i$ is
$f^{X_{i}} \simeq 1/M^{X_{i-2}X_{i-1}X_{i}X_{i+1}X_{i+2}}_{RRW}$
which agrees with Eq.(15) in Ref.\cite{Song2017proofreading}.

The above logic can be directly extended to $h-$order processes.
Under the corresponding bio-relevant conditions, one can show that the $(2h+1)$-neighbors $X_{i-h}...X_{i-1}(X_{i})X_{i+1}...X_{i+h}$ contribute overwhelmingly to $\Pi^{X_{i-h}...X_{i}...X_{L}}_{\ \alpha_{i-h}...\alpha_{i}}$.
With this generalized NN approximation, we can readily calculate any positional quantities at $i$ by assuming $g^{X_{i}...X_{i+h-1}...X_{L}}_{\ \alpha_{i}...\alpha_{i+h-1}} \simeq k^{X_{i}...X_{i+h-1}X_{i+h}...X_{L}}_{\ \alpha_{i}...\alpha_{i+h-1}R}$.

\section{Summary}
Studies on the template-specific fidelity of DNAPs are important to understand how genetic mutations are generated and controlled.
While biochemical experiments have offered much insights on this issue, systematic
theoretical studies are still rare.
The only work appeared two years ago\cite{gaspard2017Iterated}
, which dealt with the long-time limit of the replication kinetics and proposed an iteration algorithm to numerically compute the fidelity or velocity profile.
In this manuscript, we proposed a different method, based on the first-passage description of the replication process, to address these problem for complicated processes with high-order neighbor effects.
Although the boundary conditions in our method is different from the periodic boundary condition in the iteration algorithm, it was verified numerically that these two choices always give the same results.

Our method, however, largely simplifies the calculations by introducing a closed set of kinetic equations and is more convenient for approximate analytical calculations. We showed that the positional fidelity and velocity can be reliably estimated by the nearest-neighbor approximations under bio-relevant conditions.
The analytical expressions of the positional fidelity were derived, which shows intuitively that how the template-dependent proofreading pathways could be coordinated with the polymerization pathways to achieve high fidelity.
These results also indicates that the positional quantities are only dependent on the closely surrounding template sequence and irrelevant to the statistical features (e.g, long-range correlations) of the template sequence, which is consistent with the widely-held belief that replication mutations are randomly distributed among genome.
This is also a justification of the somewhat arbitrary choices of the template sequence (e.g, any expanded sequence containing the sequence under study can be chosen as the template) and the initial condition (at the reflecting boundary) in our method.

Our method can also be applied to more realistic cases in which either the addition or the deletion of monomers consists of multiple sub-steps (e.g, see Ref.\cite{tsai2006new}).
The basic theory in Sec.\ref{basic} can be slightly modified to handle these problems without appealing for additional steady-state assumptions which are usually adopted to model multi-step processes in biochemistry (e.g, the well known Michaelis-Menten kinetics which is also used to study the DNA replication in Ref.\cite{gaspard2016kinetics1}).
Comprehensive discussions on this issue, as well as applications to real DNA replication systems, will be presented elsewhere.

\section{Acknowledgments}
The authors thank the financial support by National Natural Science Foundation of China
(No.11675180, No.11574329, No.11774358), Key Research Program of Frontier Sciences of CAS (No. Y7Y1472Y61), the CAS Biophysics Interdisciplinary Innovation Team Project (No.2060299), CAS Strategic Priority Research Program (No. XDA17010504)£¬and the Joint NSFC-ISF Research Program(No. 51561145002)¡£

\appendix
\appendixpage
\addappheadtotoc

\begin{figure*}
\includegraphics[width=12cm]{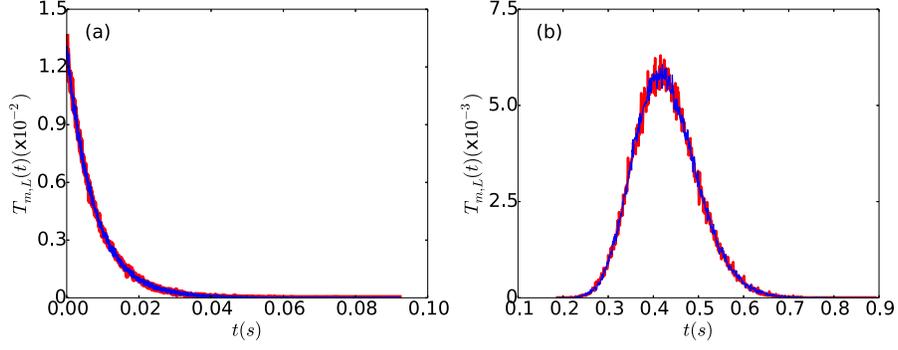}
\caption{Verification of Eq.\ref{dwelltime} by simulations for the random template. The statistics are made over $10^5$ simulations under Parameter 1.  \textit{Red}: dwell time distribution reconstructed from simulations of the original FP process (from the template position 1 to L). \textit{Blue}: dwell time distribution reconstructed according to Eq.\ref{dwelltime}, from simulations of the new FP processes (from the position $m$ to L with $a$ or $b$ at $m$).  (a)$T_{99,100}(t)$; (b)$T_{51,100}(t)$.}
\label{f6}
\end{figure*}

\subsection{The dwell time distribution at position $m$ } \label{dwelltimedis}
It's mentioned in the main text that $\Gamma_{m}$ is the mean dwell time of the growing chain at template position $m$, according to its definition $\Gamma_{m}=\int^{+\infty}_{0}p^{X_m}(t)dt$. To understand this, one can imagine $N$ simulation trajectories generated by Gillespie algorithm (the first-passage process is divided into infinitesimal intervals $dt$ )and select the $N_m(t)$ trajectories in which the growing end stays at position $m$ at time $t$, i.e, to get a statistics of $p^{X_m}(t)= N_m(t)/N$ as well as the infinitesimal dwell time $dt$ at $m$.
As the copolymerization proceeds, the total dwell time at $m$ contributed from all the $N$ trajectories should be $\int^{+\infty}_{0}N_m(t)dt$, hence the mean dwell time per trajectory is given by $\int^{+\infty}_{0}p^{X_m}(t)dt$.

One can further investigate the dwell time distribution at $m$.  Denote the total dwell time the growing chain spends when its length $n$ reaches the region $m\leq n < L$ as $t$, and define the corresponding time distribution as $T_{m,L}(t)$, we have $\int^{\infty}_{0}T_{m,L}(t)dt=1$ and it's also known from above that $\int^{\infty}_{0}tT_{m,L}(t)dt=\sum_{n=m}^{L}\Gamma_{n}$.
From the simulation results, we found that $T_{m,L}(t)$ can be precisely expressed as
\begin{eqnarray}\label{dwelltime}
T_{m,L}(t)=\sum\limits_{\alpha_{m}=a,b}P^{X_{m}}_{\ \alpha_{m}}T^{\alpha_{m}}_{m,L}(t)
\end{eqnarray}

$P^{X_{m}}_{\ \alpha_{m}}$ is the final probability distribution at $m$, as calculated in the main text.
$T^{\alpha_{m}}_{m,L}(t)$ is defined as the first-passage time distribution of a new imaginary replication process which initiates at the template position $m$ with initial conditions $q^{X_{m}}_{\ \alpha_{m}}=1$ ($\alpha_{m}=a$ or $b$) and again terminates at position $L$. This equation can be precisely verified by numerical calculations(Fig.\ref{f6}).

One can also calculate the positional dwell time distribution $T_{m}(t)$, by using the convolution relation
$$T_{m,L}(t)=\int_{0}^{t}T_{m}(\tau)T_{m+1,L}(t-\tau)d\tau$$
with
\begin{eqnarray}\label{}
T_{L-1}(t)& \equiv &T_{L-1,L}(t)\nonumber \\
&=& \sum\limits_{\alpha_{L-1}=a,b}P^{X_{L-1}}_{\ \alpha_{L-1}}(k^{X_{L-1}X_{L}}_{\ \alpha_{L-1}a}+k^{X_{L-1}X_{L}}_{\ \alpha_{L-1}b})\nonumber\\
&&\cdot e^{-(k^{X_{L-1}X_{L}}_{\ \alpha_{L-1}a}+k^{X_{L-1}X_{L}}_{\ \alpha_{L-1}b})t}\nonumber
\end{eqnarray}

\subsection{The eigenvector approximation}\label{eigenapp}
Aperiodic and irreducible stochastic matrices like $M$ have an important property according to Perron-Frobenius theorem, i.e, their largest eigenvalue is $\lambda_1 = 1 $ which always associates with one and only one positive eigenvector $V_1$ being properly normalized to 1.
Other eigenvalues and eigenvectors are denoted as $\lambda_i$ and $V_i, i=2,3,...n$, $n$ is the dimension of the matrix.
For stochastic matrices like $M$ under bio-relevant conditions, $\lambda_i (i\geq 2)$ are always far less than 1.
Any probability distribution vector $P$ can be decomposed as $P = V_1 + \sum_{i>1} s_iV_i$, so
$PM = V_1 + \sum_{i>1} \lambda_is_iV_i$.
If $PM$ differs not much from $P$, the second summation in the above equality is always far less than $V_1$, so $PM$ can be approximated by $V_1$.

On the other hand, we also know that $ (P^{X_i}_{\ R}, P^{X_i}_{\ W})= (P^{X_{i-1}}_{\ R}, P^{X_{i-1}}_{\ W}) \cdot M^{X_{i-1}{X_i}{X_i+1}}$, and $(P^{X_i}_{\ R}, P^{X_i}_{\ W})$ is indeed not too different from $(P^{X_{i-1}}_{\ R}, P^{X_{i-1}}_{\ W})$ (they both are around (1,0)).
So we can safely approximate $(P^{X_i}_{\ R}, P^{X_i}_{\ W})$ by the eigenvector $V_1$ of the matrix $M^{X_{i-1}{X_i}{X_i+1}}$.

\subsection{The template sequences and kinetic parameters }\label{temppara}
The DNA template and kinetic parameters used in the numerical computations and simulations in the main text are shown in Table \ref{ranseq} and Table \ref{paralist}.

\begin{table*}[!tpb]
\caption{Random template}
    \label{Tab:01}
    \begin{tabular}{|ccccc|}
    \hline
    1-10&11-20&21-30&31-40&41-50\\
    BAABAAABBB&AAAAABABAA&BBBBAAABBB&ABBABBAAAB&BBABAABAAA\\
    \hline
    51-60&61-70&71-80&81-90&91-100\\
    BBBBBBABAA&ABBBBABABB&AAAABAABBB&ABBBBBBBBA&AABABABABB\\
    \hline
    \end{tabular}
    \label{ranseq}
\end{table*}

\begin{table*}[!tpb]
\caption{kinetic parameters($s^{-1}$, simulation time unit)}
    \label{Tab:03}
\begin{tabular}{c|cccc|cccc|cccc}
\hline
Parameters & \multicolumn{4}{|c|}{1}&\multicolumn{4}{|c|}{2}&\multicolumn{4}{|c}{3}\\
\hline
Template& & & & & & & & & & & & \\
 & AA&AB&BA&BB& AA&AB&BA&BB& AA&AB&BA&BB\\
di-unit& & & & & & & & & & & & \\
\hline
 $k_{aa}$	&	65.0	&	45.0	&	76.0	&	45.0	&	250.0	&	0.42	&	0.52	&	0.0001  &12344.0 &55325.0 &43.0 &5436.0 \\
    $k_{ab}$	&	68.0	&	45.0	&	64.0	&	97.0	&	0.77	&	200.0	&	0.0001	&	0.8     &	34.0	&	6325.0	&	2456.0	&	54.0 \\
    $k_{ba}$	&	54.0	&	95.0	&	56.0	&	78.0	&	0.14	&	0.0001	&	150.0	&	0.56    &	3432.0	&	342.0	&	243.0	&	5456.0 \\
    $k_{bb}$	&	45.0	&	66.0	&	80.0	&	67.0	&	0.0001	&	0.92	&	0.69	&	300.0   &	657890.0	&	3424.0	&	54.0	&	1324.0 \\
    \cline{1-13}
    $r_{aa}$	&	12.0	&	23.0	&	7.0	&	4.0	&	0.0065	&	0.018	&	0.026	&	2.0     &    314.0  &	3244.0	&	543.0	&	32.0 \\
    $r_{ab}$	&	16.0	&	24.0	&	16.0	&	4.0	&	0.033	&	0.0007	&	3.0	    &	0.011   &  	2.0	   &	3.0	    & 	434.0	&	2.0 \\
    $r_{ba}$	&	22.0	&	9.0	&	17.0	&	28.0	&	0.036	&	5.0	    &	0.0018	&	0.067   & 	3.0	   &	4.0	    &	543.0	&	234.0 \\
    $r_{bb}$	&	14.0	&	23.0	&	12.0	&	19.0	&	2.0	    &	0.046	&	0.098	&	0.0015  &	43.0   &	5.0	    &	73.0	&	12.0 \\
\hline
\end{tabular}
    \begin{flushleft}
    Parameters 1: the addition rates and deletions rates are of the same orders of magnitude, which is different from the bio-relevant conditions. \\
    Parameters 2: bio-relevant conditions in which $R$ and $W$ (base pairs) can be uniquely defined for each template unit (say $A$-$a$, $B$-$b$). \\
    Parameters 3: all the rates are randomly assigned, which strongly violates the bio-relevant conditions: no $R$ or $W$ can be properly defined for each template unit.
  \end{flushleft}
  \label{paralist}
\end{table*}

\begin{table*}[!tpb]
\caption{Markov template}
    \label{Tab:02}
    \begin{tabular}{|ccccc|}
    \hline
    1-10&11-20&21-30&31-40&41-50\\
    AAAAAAAAAA&AAAAAAAAAA&BBBBBBBBBB&BBBBBBBBBB&BBBBBBBBBA\\	
    \hline
    51-60&61-70&71-80&81-90&91-100\\
    AAAAABBAAA&AAAAAAAAAA&AAABBBBBBB&BBBBBBBBBB&BBBAAAAAAA\\
    \hline
    \end{tabular}
    \label{marseq}
\end{table*}

In Sec.\ref{NNapp}, it has been shown that our first-passage approach and nearest-neighbor approximation can reliably reproduce the fidelity and velocity profile under bio-relevant conditions, which means that these positional quantities are irrelevant to the long-rang properties of the template sequence.
To better illustrate this, we have carried out numerical computations for a Markov chain template sequence (Table \ref{marseq}) in which the probability of consecutive $A$s (or $B$s) is taken as 0.8.  Our results (see Supplementary Materials) clearly show that the NN approximation still holds for such strongly-correlated template sequences.

\end{document}